\documentclass[aps,reprint,nofootinbib]{revtex4-1}
\usepackage{graphicx,amsfonts,amsmath,amssymb}
\usepackage{bm}
\usepackage{url}

\makeatletter
\g@addto@macro{\UrlBreaks}{\UrlOrds}
\makeatother

\begin{document}
\title{A Two-Page Derivation of Schr\"odinger's Equation}
\date{\today}
\author{C. Baumgarten}
\affiliation{Switzerland}
\email{christian-baumgarten@gmx.net}

\def\begeq{\begin{equation}}
\def\endeq{\end{equation}}
\def\bquo{\begin{quotation}}
\def\equo{\end{quotation}}
\def\begary{\begeq\begin{array}}
\def\endary{\end{array}\endeq}
\newcommand{\myarray}[1]{\begin{equation}\begin{split}#1\end{split}\end{equation}}
\def\bmtx{\left(\begin{array}}
\def\emtx{\end{array}\right)}
\def\d{\partial}
\def\h{\eta}
\def\w{\omega}
\def\W{\Omega}
\def\s{\sigma}
\def\eps{\varepsilon}
\def\e{\epsilon}
\def\a{\alpha}
\def\b{\beta}
\def\g{\gamma}
\def\y{\gamma}
\def\d{\partial}
\def\S{{\Sigma}}

\def\leftD#1{\overset{\leftarrow}{#1}}
\def\rightD#1{\overset{\rightarrow}{#1}}

\begin{abstract}
We give an exceptionally short derivation of Schr\"odinger's equation by 
replacing the idealization of a point particle by a density distribution.
\end{abstract}
\maketitle

\section{Introduction}

\begin{quotation}
There is strong reason to believe that Nature works according to mathematical laws.
All the substantial progress of science supports this view.\\
\mbox{}\hfill{\it -- P.A.M. Dirac}~\cite{Dirac}
\end{quotation}

Contrary to a common textbook view that the Schr\"odinger equation (SEQ) can not possibly
be derived~\cite{FLS,Messiah}, the number of published papers that either present 
derivations of the SEQ and/or (semi-) classical perspectives on the SEQ, increases
continuously\footnote{See Refs.~\cite{Madelung1927,Fuerth1933,Fenyes1948,Fenyes1952,Weizel1953,Kershaw1964},
~\cite{Froehlich1973,Bess1973,Bess1974,Baublitz1988,Derbes1996,Andrews1997,Briggs2001},
~\cite{Bergeron2001,Kaniadakis2002,Fritsche2003,Gutzwiller2003a,Gutzwiller2003b},
~\cite{Gutzwiller2003c,Groessing2004,Pesci2005,Skala2005,Scarfone2005,Parwani2006},
~\cite{Brenig2007,Parwani2007,Scarfone2007,Nottale2007,PenaAuerbach2008,Davidson2008},
~\cite{Klein2009,Ogiba2011,Chiribella2011,Budiyono2012a,Budiyono2012b,Budiyono2013},
~\cite{SGKS2013,Hofmann2014,Tessarotto2016,Franca2016,Budiyono2017},
~\cite{Katsnelson2021,Claeys2021,Gao2021,Gaukhman2022},﻿
~\cite{Grinwald2022,Lima2022}. This list, though rather long, is certainly
still incomplete.}.
We shall present yet another derivation of the SEQ based mostly on formal
mathematical considerations.

Classical mechanics is often identified with
the mechanics of ``point particles'' or ``mass points''. There is no classical
theory of extended continuous microscopic objects, despite the fact that 
``natural philosophy'' defined material objects as {\it res extensa}.
But since classical physics identifies matter by it's position in
space and time, it seems unclear how to assign a common identity to the
volume elements of extended objects. Newton tried to circumvent the problem
by the involvement of a supernatural power~\cite{NewtonOpticks}:
\bquo
    [...], it seems probable to me, that God in the
    Beginning form'd Matter in solid, massy, hard,
    impenetrable, movable Particles, of such Sizes
    and Figures, and with such other Properties, and
    in such Proportion to Space, as most conduced to
    the End for which he form'd them; and that these
    primitive Particles being Solids, are incomparably 
    harder than any porous Bodies compounded of them;
    even so very hard, as never to wear or break in pieces;
    no ordinary Power being able to divide what God
    himself made one in the first Creation.
\equo
Boscovich~\cite{Boscovich}, an influential Serbian Jesuit scientist, suggested
a different approach by the invention of point particles in the strict
mathematical sense~\cite{Glazebrook}:
\bquo
According to Boscovich an atom is an indivisible point, having position
in space, capable of motion, and possessing mass. [...]
It has no parts or dimensions: it is a mere geometrical point without extension 
in space: it has not the property of impenetrability, for two atoms can, it is 
supposed, exist at the same point.
\equo

But though Boscovich' position seems to have prevailed in modern textbooks
on analytical mechanics, a survey of textbooks on mechanics of the 19th
century, before the advent of quantum theory, provides evidence that the
``classical'' point particle never was accepted as an undisputable
and self-consistent idea~\cite{vlh_paper}. Also James Clerk Maxwell,
to give just one example, disagreed with Boscovich~\cite{Glazebrook}:
\bquo
We make no assumption with respect to the nature of the small parts -- 
whether they are all of one magnitude. We do not even assume them to
have extension and figure. Each of them must be measured by its mass, 
and any two of them must, like visible bodies, have the power of acting on one
another when they come near enough to do so. The properties of the body or medium
are determined by the configuration of its parts.
\equo

Rohrlich critically reviewed the idealization of the classical point
particle~\cite{Rohrlich}. He wrote that 
``in the point limit, classical physics cannot be expected to make sense at all''. 
As he explains, a point charge, regarded from the standpoint of classical 
electrodynamics, implies infinite self-energy. Hence, according 
to Rohrlich, ``the concept of a "classical point particle" is, in view of 
quantum mechanics, an oxymoron. Quantum mechanics tells us that below a 
certain magnitude of distance, usually characterized by a Compton wavelength,
classical physics ceases to be reliable; predictions made by classical
mechanics or classical electrodynamics must be replaced by quantum
mechanical predictions.''

As our derivation will demonstrate, the SEQ implicitely suggests yet
another solution to the problem of extended objects, namely it transforms
the problem of the ``parts'' of extended objects into Fourier space:
The extended particle is then not an ensemble of ``material points''
in space, but an ensemble of waves in Fourier space.
The question whether we interpret the ``extended object'' as a probability
distribution or directly as a matter density, however, is not part
of the following derivation.

\section{The Schr\"odinger Equation}

Consider that an extended particle (or the probability to find a point
particle, respectively), is represented by a normalizable spatial 
density distribution $\rho(t,\vec x)$
\begeq
\int\,\rho(t,\vec x)\,d^3x=1\,.
\label{eq_norm}
\endeq
We regard the density as a positive definite quantity: $\rho\ge 0$. In order
to fulfill this requirement, we express the density by the square modulus 
of (a sum of) auxiliary functions $\psi_i(t,\vec x)$ such that
\begeq
\rho(t,\vec x)=\sum_i\,\psi_i^2(t,\vec x)\,.
\endeq
For simplicity - and without loss of generality - we may use complex numbers
and chose the following positive semidefinite expression
\begeq
\rho(t,\vec x)=\psi(t,\vec x)^\star\psi(t,\vec x)\,.
\endeq
The auxiliary function $\psi(t,\vec x)$ is then due to Eq.~\ref{eq_norm} square 
integrable. It is therefore a member of some Hilbert space and the Fourier
transform $\psi(t,\vec k)$ is known to exist:
\begeq
\psi(t,\vec x)\propto\,\int\,\tilde\psi(\w,\vec k)\,\exp{[-i\,(\w\,t-\vec k\cdot\vec x)]}\,d^3k\,d\w.
\endeq
But the Fourier transformation alone, without any further constraint, does not
yield a physical model of anything. All known {\it physical} waves are 
characterized by a relation between frequency and wavelength, i.e. by a
dispersion relation. It can be demonstrated that the necessary existence of a dispersion
relation is a consequence of a causality requirement~\cite{Toll,Guettinger,Dethe2019}.
The dispersion relation then allows to obtain an expression for the velocity
of the ``wave packet'', namely the so-called {\it group velocity}~\cite{Hamilton,Brillouin,Whitham}:
\begeq
\vec v_{gr}=\vec\nabla_k\,\w(\vec k)=({\d\w\over\d k_x},{\d\w\over\d k_y},{\d\w\over\d k_z})^T\,,
\label{eq_dispersion}
\endeq
where $\w(\vec k)$ is the mentioned dispersion relation.
Eq.~\ref{eq_dispersion} has precisely the form of the velocity equation of 
classical Hamiltonian mechanics, which relates the velocity to the 
gradient of the energy (i.e. the Hamiltonian function) in momentum space:
\begeq
\vec v=\vec\nabla_p\,{\cal H}(\vec p)
\endeq
Hence, if the wave packet is supposed to provide a description of a classical
particle, the (average) velocity of the wave packet must agree with the
Hamiltonian expression~\footnote{Several authors of standard textbooks on QM, 
for instance Messiah~\cite{Messiah}, Schiff~\cite{Schiff} as well as 
Weinberg~\cite{Weinberg} use this equation, not to derive Schr\"odinger's
equation, but merely to make it ``plausible''.}:
\begeq
\vec\nabla_k\,\w(\vec k)=\vec\nabla_p\,{\cal H}(\vec p)\,.
\label{eq_base}
\endeq
A solution that complies with Eq.~\ref{eq_base}, where energy and momentum 
have the usual units, requires the introduction of a proportionality constant 
with the unit of action, let's call it $\hbar$. It is, in the context of our
derivation, {\it not} a substantial physical unit and does not imply any
quantization of energy. Conversion factors can not be derived theoretically~\cite{Weyl1930}:
\bquo
The constants $c$ and $h$, the velocity of light and the quantum of action,
have caused some trouble. The insight into the significance of these
constants, obtained by the theory of relativity on the one hand and quantum
theory on the other, is most forcibly expressed by the fact that they do not
occur in the laws of Nature in a thoroughly systematic development of these
theories.
\equo
$\hbar$ is merely a conversion factor that is necessary
due to the contingent historical choice of units. The quantization of atomic
energy levels is, from a logical point of view, an ``output of the theory'',
not an input~\cite{Tong}. It is entirely due to boundary conditions as a
free particle can have any (kinetic) energy.

Since Eq.~\ref{eq_base} equates
two derivatives, one has to allow for ``integration constants'', i.e. functions 
(potentials) that do not depend on momentum or wave-number. We therefore
use 
\myarray{
\hbar\,\vec k&=\vec p+q\,\vec A(\vec x)\\
\hbar\,\w(\vec k)&={\cal H}(\vec p)+q\,\phi(\vec x)\\ 
}
with some arbitrary constant $q$. Note that this is not a physical hypothesis, 
but a formulation of the linear restriction that the wave ensemble has to comply 
with, if it is supposed to consistently represent a classical particle in 
{\it some} way. The ``integration constants'' $\phi$ and $\vec A$ are well 
known in classical mechanics and can assumed to be zero for a free particle.

The total normalization must of course be preserved and this requires that the
wave motion is {\it adiabatic}. Max Born referred to the classical adiabatic
invariance of the 
phase space volume $\Phi=\mathrm{const}$ and to the fact that energy (change) 
and frequency (change) are, in such processes, proportional to each 
other $\delta{\cal E}=\Phi\,\delta\w$~\cite{Born}. Furthermore it is long
known that the real and imaginary components of the wave function 
are subject to Hamiltonian motion~\cite{Strocchi,Ralston1989}. 

Instead of showing that Schr\"odinger's equation implies Hamiltonian notions,
we consider the reverse argument: if one presumes the validity of energy
conservation and hence of classical Hamiltonian notions in wave dynamics, 
then Eq.~\ref{eq_base} is automatically valid. The de-Broglie relations
then immediately follow:
\myarray{
{\cal E}&=\hbar\,\w\\
\vec p&=\hbar\,\vec k\,.
\label{eq_deBroglie}
}
Inserting this into the Fourier transform gives:
\begeq
\psi(t,\vec r)\propto\int\,\tilde\psi({\cal E},\vec
k)\,\exp{[-i\,({\cal E}\,t-\vec p\cdot\vec x)/\hbar]}\,d^3p\,d{\cal E}\,.
\label{eq_ft1}
\endeq
Then the energy is equal to the time derivative, and the momentum to the
spatial gradient. That is, the canonical ``quantization''
rules directly follow: 
\myarray{
{\cal E}\,\psi(t,\vec r)&=i\,\hbar{\d\over\d t}\psi(t,\vec r)\\
{\vec p}\,\psi(t,\vec r)&=-i\,\hbar\,\vec\nabla\,\psi(t,\vec r)\,.
}
Using these relations one can express the classical (kinetic) energy of a free particle 
${\cal E}={\vec p^2\over 2\,m}$, which results in Schr\"odinger's equation for a free particle:
\begeq
i\,\hbar{\d\over\d t}\psi(t,\vec x)=-{\hbar^2\over 2\,m}\,\vec\nabla^2\,\psi(t,\vec r)\,.
\endeq
Adding a potential energy (density) $\rho(t,\vec x)\,\phi(\vec x)$ readily yields Schr\"odingers
equation for a particle in potential $\phi(\vec x)$:
\begeq
i\,\hbar{\d\over\d t}\psi(t,\vec x)=\left(-{\hbar^2\over 2\,m}\,\vec\nabla^2+\phi(\vec x)\right)\,\psi(t,\vec r)\,.
\endeq
This derivation of Schr\"odinger's equation is short and rigorously follows
from the assumption that the presence of a material particle must have a
mathematical representation by a finite normalizable positive semidefinite density.
This could be described as the logical classical alternative to the representation
of a particle as mathematical point without extension.
It provides a new perspective on the relationship between classical mechanics
and quantum theory and shows that, contrary to usual assertions, these theories 
are not mathematically disjunct.

Our derivation neither presumed nor suggested a specific interpretation
of the ``wave function''. It only provides a different perspective on the
mathematical form of Schr\"odinger's equation and demonstrates that
there are indeed classical arguments which lead to an action constant
of universal physical significance.

\section{Summary and Conclusions}

As Rohrlich's analysis reveals, the alleged intuitiveness and logic of the 
notion of the point particle fails, on closer inspection, to provide a
physically and logically consistent {\it classical} picture. If we dispense
this notion, Schr\"odinger's equation can be derived and might be regarded 
as a kind of {\it regularization}, which allows to circumvent
the problematic infinities of the point-particle-idealization.
Nonetheless it is ahistorical to depict the point-particle-idealization
as ``classical'', since most textbook authors of the pre-quantum era stayed,
like Maxwell, agnostic about the true nature of the ``smallest parts''~\cite{vlh_paper}.

The given presentation uses the ``Born rule'', which states that
$\psi^\star\,\psi$ is a (probability) density, as a mathematical method
rather than as a part of it's interpretation. 

However, as well-known (though often ignored), Schr\"odinger's equation
is not the most fundamental equation of quantum theory. It has to be
derived from the Dirac equation.
Only the Lorentz covariant Dirac equation provides full compatibility with
electromagnetic theory. It was shown elsewhere how the Dirac equation can,
in momentum space, be derived from Hamiltonian methods~\cite{qed_paper,osc_paper,uqm_paper}.
The derivation automatically yields the Lorentz transformations, the Lorentz
force law~\cite{rdm_paper,geo_paper,lt_paper} and Maxwell's
equations~\cite{qed_paper} in a single coherent framework.

\bibliography{seq_paper}{}
\bibliographystyle{unsrt}

\end{document}